# The atmospheric global electric circuit: An overview

Devendraa Siingh[a,b*], V. Gopalakrishnan[a], R. P. Singh[c], A. K. Kamra[a], Shubha Singh[c], Vimlesh Pant[a], R. Singh[d], and A. K. Singh[e]

[a]Indian Institute of Tropical Meteorology,  Pune-411 008, India
[b]Institute of Environmental Physics, University of Tartu, 18, Ulikooli Street,
   Tartu- 50090, Estonia
[c]Department of Physics, Banaras Hindu University, Varanasi-221 005, India
[d]Indian Institute of Geomagnetism,  Mumbai-410 218, India
[e]Physics Department, Bundelkhand University, Jhansi, India

**Abstract:**
Research work in the area of the Global Electric Circuit (GEC) has rapidly expanded in recent years mainly through observations of lightning from satellites and ground-based networks and observations of optical emissions between cloud and ionosphere. After reviewing this progress, we critically examine the role of various generators of the currents flowing in the lower and upper atmosphere and supplying currents to the GEC. The role of aerosols and cosmic rays in controlling the GEC and linkage between climate, solar-terrestrial relationships and the GEC has been briefly discussed. Some unsolved problems in this area are reported for future investigations.

-------------------------------------------------------------------------------------------------------------

*Address for corrspondence
Institute of Environmental Physics, University of Tartu, 18, Ulikooli Street
Tartu-50090, Estonia
e-mail; devendraasiingh@tropmet.res.in
         amartyadsiingh@yahoo.com
Fax     +37 27 37 55 56

# 1. Introduction

The global electric circuit (GEC) links the electric field and current flowing in the lower atmosphere, ionosphere and magnetosphere forming a giant spherical condenser (Lakhina, 1993; Bering III, 1995; Bering III et al., 1998; Rycroft et al., 2000; Siingh et al., 2005), which is charged by the thunderstorms to a potential of several hundred thousand volts (Roble and Tzur, 1986) and drives vertical current through the atmosphere's columnar resistance. The current causes weak electrification of stratified clouds (Harrison and Carslaw, 2003) and produces a vertical potential gradient in atmospheric layers near the surface. Horizontal current flows freely along the highly conducting earth's surface and in the ionosphere, which is closed by the current flowing from the ground in to the thunderstorm and from the top of the thunderstorm to the ionosphere and back from the ionosphere to the ground through the global fair-weather load resistance (~ 100 Ω). Different regions of the atmosphere including the ionosphere and magnetosphere are electro-magnetically linked (Singh et al., 2004; Siingh et al, 2005). In an active thunderstorm the upward current flows from thunderstorm to the ionosphere, which is known as Wilson current. This current spreads around the globe through the ionosphere/magnetosphere along the geomagnetic field lines to the opposite hemisphere. Figure 1 shows the global distribution of the current (Roble and Tzur, 1986) in which the current returns to the surface of the earth as the fair-weather air-earth current. This figure also shows the location of relevant layers of the atmosphere which is divided into four coupled regions i.e. troposphere, middle atmosphere, ionosphere and magnetosphere. The orography of the earth is also considered (Roble and Tzur, 1986). In this diagram the ionosphere and magnetosphere are treated as the passive elements of the circuit. The upward current from the earth's surface to the bottom of the thundercloud consists of field dependent current, convection current, lightning current, precipitation current and displacement current (Roble, 1991). Any perturbation in the interplanetary or atmospheric environment causes a variation in electrical conductivity and hence variation in current/electric field system of the atmosphere. The variations on solar surface causes variations in the solar wind parameters, which can be coupled with the stratosphere and troposphere leading to modulation of current density in the global atmospheric electric circuit from the ionosphere to the earth. The small-scale chaotic variation of conductivity in the mixing layer make it difficult to measure the systematic variation of vertical current due to solar wind input. Even very small changes (1~ 3%) in the cosmic ray flux in the equatorial regions due to variation in solar wind inputs may affect the thunderstorm charging current and ionospheric potential.

Recent observations of optical emissions above the top of the thunderstorm show that the electrical behaviour of the region is quite different than it was assumed in earlier models of GEC (Singh et al., 2004; Siingh et al., 2005). Rycroft (2006) has discussed tele-connections between thunderstorms, lightning and optical emissions above thunderstorms. At mid latitudes precipitated burst of energetic electrons may lower the ionospheric equi-potential surface above the thundercloud sufficiently to trigger an upward lightning discharge to the ionosphere (Rycroft, 1994). The flux density of precipitated electrons depends upon solar and geomagnetic conditions. In fact solar wind, solar flares, galactic cosmic rays, ionospheric-magnetospheric dynamo, thunder cloud, geomagnetic disturbances, solar magnetic sector boundary crossings, solar cycle variations, auroral activity etc affect the components of GEC (Lakhina 1993; Tinsley, 2000; Singh et al., 2004). Solar wind and geomagnetic storm affect fair-weather current via aerosol distributions/ cloud microphysics (Pudovkin and Babushkina, 1992; Tinsely, 2000; Carslaw et al, 2002; Harrison and Carslaw, 2003; Tinsley and Yu, 2003) by changing the pressure/ temperature distribution of the troposphere or by changing its dynamics. The earth's climate and climatic changes have direct connection with lightning activity (Williams, 2005), which has direct linkage with GEC. Thus, GEC may constitute a variable physical mechanism linking space weather and the earth's weather and climate (Rycroft et al, 2000; Rycroft and Fullekrug, 2004).

In the above linkages, there are various physical processes having relaxation time lying between microseconds to hours, which require further study to improve our understanding, particularly electrical processes in the earth's environment. In the recent past a number of programs have been launched to understand some of the related phenomena/ processes; such as SPECIAL (Space Processes and Electrical Changes Influencing Atmospheric Layers) organized by European Space Agency to study (i) global atmospheric electric circuit, (ii) charge particle fluxes, events and statistics and

(iii) sprites and lightning (Rycroft and Fullekrug, 2004). Other programs include STEP (Solar Terrestrial Energy Program), GEM (Geospace Environment Modeling), GAEM (Global Atmospheric Electrical Measurements) and IGBP (International Geospace-Biosphere Program). Thus, several working groups have been engaged to study sun-weather relationship and GEC.

In the recent past, many review articles have appeared describing various specific aspects of the GEC (e.g. Roble and Tzur, 1986; Bering III, 1995; Bering III et al., 1998; Rycroft et al., 2000; Tinsley, 2000; Singh et al. 2004; Harrison, 2005, Siingh et al., 2005; Williams, 2005; Rycroft, 2006). Recent developments and studies in this area, however, point out the need of having a comprehensive view of the flow of currents in different regions of the Earth's environment and possible linkage of the GEC with several other phenomena such as cosmic rays, atmospheric aerosols, weather and climate, sprites, blue jets, elves etc. In this paper, we briefly review the present status of GEC and highlight the possible linkages with other phenomena relating it to weather and climate. Section 2 briefly discusses thunderstorms, optical emissions, Schumann resonances, ionospheric dynamo and magnetospheric dynamo, which constitute the major sources of GEC. In section 3, we discuss various models of GEC proposed from time to time. It has been suggested that the mathematical modeling of GEC should include the intense current between the top of the thunderstorm and the ionosphere during optical emissions, which was not included earlier. In section 4, we discuss the electrical conductivity and columnar resistance component of GEC, which controls the magnitude of electric potential and current. The role of solar and cosmic ray variability is discussed briefly. The variation of global temperature (climate change) and its relation with global atmospheric electric circuit is discussed in section 5. It is noted that even 1% increase in global surface temperature could result into a 20% increase in ionospheric potential (Price, 1993). Aerosols act as a mediator of cloud microphysics, precipitation, cloud electrification and lightning. This aspect has been briefly discussed in section 6, where as section 7, is devoted to cosmic rays, climate and global electric circuit. The importance of GEC study to our society is presented in section 8. Finally, some recommendations for future work are presented in section 9.

## 2. Generators and sources of the GEC

The main sources of electric fields and currents in the GEC are thunderstorms, in the troposphere and dynamo situated in the ionosphere and magnetosphere. The current output from the thunderstorms around the globe maintains a vertical potential difference ~ 300 kV between the ground and the ionosphere (where the later is at positive potential), with a current flow of about $10^3$ A. The dynamo in the ionosphere is produced by tides generated in situ and tides propagating upward from the lower atmosphere. These tides generate horizontal potential differences of 5-15 kV with the current flow of ~ $10^5$ A within the ionosphere. On the other hand, the magnetospheric dynamo is driven by the interaction of solar wind with the Earth's geomagnetic field and generates a horizontal dawn-to-dusk potential drop of ~ 40 - 100 kV across the magnetic conjugate polar cap. Measurements have never shown a complete absence of fair weather electric field for any length of time, thereby suggesting continuous operation of thunderstorm and other generators in maintaining the current flowing in the global circuit. The role of different generators in generating the electric field/current in different regions is discussed below.

### 2.1 Thunderstorm

The major source of dc electric field is the thunderstorm. Below the cloud, a net negative charge is transferred from the thundercloud to the earth and above it positive charge is transferred to the conducting upper atmosphere/ionosphere making it at positive potential. Charge separation inside thunderstorms leads to the development of huge electric potentials and associated energized charge particle beam, which is known to radiate wide spectrum of electromagnetic waves such as optical emissions, X-rays and gamma rays (Fishman et al., 1994; Rodger, 1999; Milikh and Valdivia,1999).

Lightning activity is mainly concentrated in three distinct zones - East Asia, Central Africa and America. Lightning is more prevalent in the northern hemisphere than the southern hemisphere and mostly occurs over the land surface. The variation of lightning activity with latitude as observed from space shows that two of every three lightning flashes occur in tropical region (Williams, 1992). In addition to the tropical lightning, extra-tropical lightning activity plays a major role in the summer season in the northern hemisphere, resulting in the global lightning activity having a maximum from June to August. The worldwide thunderstorm activity as a function of universal time is shown in figure 2. In the figure, the upper panel shows annual diurnal variation of the potential gradient (V/m) on the oceans as measured by the research vessel Carnegie and Maud expeditions (Parkinson and Torrenson, 1931). The lower panel shows the annual diurnal variation of global thunderstorm activity (Whipple and Scrase, 1936). The similarity of the diurnal variations of the electric field over the oceans and of the worldwide thunderstorm activity supports the hypothesis that thunderstorms are the main electrical generators in the GEC. About 200 thunderstorms are active at any time, which are mainly concentrated over the tropical land masses during the local afternoons and cover about 10% of the earth's surface (Markson, 1978). On the remaining 90% of the earth's surface return current ~1000A (~ 1 pA/m$^2$) from the ionosphere to the earth's surface flows, which is also known as fair weather current

Ruhnke (1969) considered the thunderstorm generator as a current generator whose total Maxwell current output is independent of load, because the impedance between cloud top and the ionosphere, and between base of the cloud and ground is much larger than the load in the fair-weather regions. The load in the fair weather regions is ~300Ω, where as the resistance between cloud top and ionosphere lies in the range $10^5$ -$10^6$ Ω and between base of cloud and ground lies between $10^4$ -$10^5$ Ω (Markson, 1978). The Maxwell current density in fair-weather regions is more informative parameter than ionospheric potential because the latter depends on the columnar resistance, which exhibits complex variations (Markson, 1978). The average Maxwell current density is usually not affected by lightning discharges and varies slowly throughout the evolution of storm (Krider and Musser,1982; Pawar and Kamra, 2004). Since the Maxwell current remains steady at time when the electric field both at the ground and aloft undergoes large changes in amplitude, and some times even polarity, Krider and Musser (1982) inferred that the cloud electrification processes may be substantially independent of the electric field.

Thunderstorms also couple the troposphere to the magnetosphere through electromagnetic radiations generated from lightning discharges. The interaction of electromagnetic waves propagating in whistler mode with counter streaming energetic electrons causes precipitation of substantial fluxes of energetic electrons from the Van

Allen belts in to the atmosphere. Similarly waves propagating in left hand mode causes precipitation of energetic protons to the atmosphere. These energetic charged particles lower the ionospheric potential in the region of precipitation (Rycroft 1994). Further, the observations of the optical emissions on top of the thunderstorms and below the ionosphere show that the resistance of the region during emissions could not be as high as considered previously. Further, in the explanation of the optical emissions (sprites) relativistic breakdown mechanism is considered, during which MeV electrons are produced which are injected into the magnetosphere. Thus, situation becomes quite complex and this problem has not yet been solved. In the next section, we discuss briefly the features of optical emissions relevant to global electric circuit.

**2.1.1 Optical emission above the thunderstorm**

In the case of intense thunderstorm the return stroke current does not end in the cloud, but continues to move upward and terminates in the lower ionosphere (Lyons, 1996 and references therein). This transient current/field causes optical emissions (sprites, elves, blue jets, blue starters) in the space between the top of the cloud and the lower ionosphere. Figure 3(a) shows location of blue jets, red sprite and elves with height. In the same figure, the variation of electron density and temperature along with demarcation of troposphere, stratosphere, mesosphere and thermosphere are given. Figure 3(b) shows cloud to ground lightning, blue jets, gigantic jets sprite and elves (Neubert, 2003; Pasko, 2003). Sprites appear as cluster of short lived (~ 50 ms) pinkish red luminous columns, stretching from ~ 30 to 90 km altitude having width less than one km (Lyons, 1996; Neubert, 2003) and the maximum brightness at 66 km altitude (Wescott et al., 2001). The upper portion of the sprite is red, with wispy, faint blue tendrils extending to 40 km or lower. Boccippio et al. (1995) showed that about 80 % of sprites are associated with ELF transient events and +ve CG lightning return strokes having large peak current (> 35 kA) (Barr et al., 2000) and large $\Delta M_Q$ (total charge moment change of the thunderstorm) values. Some sprites associated with –ve CG lightning have also been observed (Barrington-Leigh et al., 1999). Sprites may occur over any area as long as energetic thunderstorms are present and they may produce detectable ELF/VLF transients (Price et al., 2002) and a vertical electric field perturbation of 0.73 V/m in stratosphere (Bering III et al. 2002). These events in early literature were popularly known as blue or green pillars and rocket like columns of optical emissions (Wood, 1951). Blue jets are slow moving fountain of blue light from the top of the cloud, where as elves are lightning induced flashes that can spread over ~ 300 km laterally around 90 km altitude in the lower ionosphere. Recently, Su et al. (2003) have observed six gigantic optical jets from the oceanic thunderstorm and each of them may remove ~30 C of charge from the ionosphere. Gigantic jets are special phenomena of the oceanic thunderstorms that establish a direct link between a thundercloud (~ 16 km altitude) and the ionosphere at 90 km elevation. ELF radio wave was detected only in four events and that no cloud-to-ground lightning was observed to trigger these events. But, observations indicate that ELF waves were generated by negative cloud–to-ionosphere discharge, which would reduce the electrical potential between the ionosphere and cloud (Rycroft, 2006). Therefore, it is necessary to modify the conventional picture of GEC to include the contribution of gigantic jet, blue jets and elves and sprites. The detailed knowledge of characteristic properties of these emissions along with frequency of events will help us to understand their contribution to the GEC (Pasko, 2003).

Cho and Rycroft (1998), using electrostatic and electromagnetic codes simulated the electric field structure from the cloud top to the ionosphere and tried to explain the observation of a single red sprite. To explain the clusters of sprites, they suggested that the positive charges are distributed in spots so that a single discharge may lead to clusters of red-sprites. The redistribution of charge and the electromagnetic pulse during lightning discharge may produce acceleration of electrons, heating and ionization of atmosphere. This may lead to strongly non-linear situation and runway electrons/electrical breakdown of the atmosphere may occur (Rycroft and Cho, 1998; Rowland, 1998). Nagano et al. (2003) evaluated the modification in electron density and collision frequency of the ionosphere by the electromagnetic pulse of the lightning discharge and explained the generation of elves. In such a situation the electrical conductivity of the atmosphere above thunderstorms enhances by about a factor of 2 from ambient values (Holzworth and Hu, 1995). These changes in conductivity could be due to thunderstorm-produced gravity waves or X-rays from lightning induced electron precipitation (Hu et el., 1989). The sprites also provide a link between tropospheric processes in thunderstorms and mesospheric processes in the upper atmosphere. Hiraki et al. (2002) suggested that sprites would change chemically the concentration of $NO_x$ and $HO_x$ in the mesosphere and lower atmosphere. These chemical changes may impact on the global cooling or heating in the middle atmosphere.

Pasko et al. (2002) have reported a video recording of a blue jet propagating upwards from a small thundercloud cell to an altitude of about 70 km. As relatively small thundercloud cells are very common in the tropics, it is probable that optical phenomena from the top of the clouds may constitute an important component of the GEC. It has motivated a reexamination of our understanding of the electrical processes and properties of the atmosphere. During a SPECIAL meeting in Frankfurt (20-23 Feb, 2003), the fundamental understanding of powering of the global electric circuit by global distribution of thunderstorms was questioned. It was argued that the power supplied by thunderstorms is insufficient to maintain a field of the magnitude observed in fair-weather regions. The inclusion of the effect of sprites and other optical phenomena also could not explain. Because, sprites etc occur in the upward branch of the global electric circuit above the thunderstorms and they are likely to influence only the upper atmosphere conductivity. Moreover, since they occur much less frequently (only one sprites out of 200 lightning) because of their association with intense lightning discharges (Singh et al., 2002), they may not play a major role in GEC (Rycroft et al., 2000). Since optical emissions could change electrical properties of the atmosphere and influence processes related with weather and climate, intense research activity in this area is required. Further, thunderstorms are also the source of Schumann resonance, which control the electrodynamics of the lower atmosphere.

**2.1.2 Schumann resonance**

Lightning discharges between thunderclouds and the ground radiate powerful radio noise burst over a wide frequency spectrum ranging from a few Hz to higher than several hundreds of Megahertz. These noises in the ELF (3 Hz -3 kHz) and VLF (3-30 kHz) range can propagate over long distances through the Earth-ionosphere waveguide (Singh et al., 2002). Especially, radio noise in the frequency range less than 50 Hz can propagate globally with extremely low attenuations (Jones, 1999) and constructive interferences of these waves result in the Earth-ionosphere cavity resonance known as Schumann resonance (SR) with their fundamental mode of eigen frequency at 8 Hz (Sentman, 1995; Huang et al., 1999; Barr et. al., 2000).

The amplitude of Schumann modes is determined by the temporal and spatial distribution of global lightning, which is intense over the tropics. The variations in solar activity or nuclear explosions produce disturbances in the ionosphere and affect SR (Schlegel and Fullekug, 1999). Solar proton events cause increase in frequency, Q-factor (i.e. band width of the resonance mode) and amplitude of the SR mode (Schlegel and Fullekrug, 1999). Sentman et al. (1996) examined the SR measurement from California and Australia during the large solar storms in the fall of 1989 and found no measurable difference in SR intensities, although they found a sudden decrease in Q-factor of the second mode, which was attributed to small changes of middle atmospheric conductivities by energetic particles. SR intensity depends upon the height of the ionosphere (Sentman and Fraser, 1999). It has solar cycle dependence (Hale, 1987) and responds to solar flares, magnetic storms (Hale, 1987) and solar proton events (Reid, 1986). However, the processes involved in it are not properly understood and efforts are being made in this direction.

Sentman (1995) has discussed the principal features of SR, which are used to monitor global lightning activity (Heckman et al., 1998; Barr et al., 2000; Rycroft el al, 2000), global variability of lightning activity (Satori, 1996; Nickolaenko et al., 1996) and sprite activity (Boccippio et al., 1995; Cummer et al., 1998; Rycroft et al., 2000). Since thunderstorm is the main source of the SR phenomenon and GEC, their link with weather and climate could be developed (Williams, 1992; Price, 1993; Price and Rind, 1994). Such links in the electromagnetic, thermodynamic, climate and climate-change characteristics of the atmosphere have greatly enhanced the interest in monitoring of electromagnetic waves and their mapping and propagation properties in different regions of the atmosphere. If lightning is the main or only source for maintenance of ionospheric potential, the measurements of SR and ionsopheric potential should produce identical results. The differences between the two results will indicate the contribution to ionospheric potential by other processes such as corona discharge from elevated objects above ground.

**2.2 Ionospheric dynamo**

The regular tidal wind system drives ionospheric plasma at dynamo layer heights and pushes it against the geomagnetic field. Ions and electrons are affected differently by these winds. While the ions being massive still move essentially with the neutrals, the geomagnetic field already controls the motion of the electrons. The differential motion of ions and electrons is responsible for horizontally flowing electric currents. Moreover, charge separation causes an electric polarization field, which is constrained by the condition of source free currents (Volland, 1987) and has been observed indirectly from backscatter measurements (Richmond, 1976). Lunar variations are usually less than 10 % of magnitude of solar variation (Matsushita, 1967). They depend not only on latitude, solar time, season and solar cycle, but also on lunar phase. Global analyses of geomagnetic lunar effects have also found significant longitudinal variations. The seasonal variations of the lunar magnetic perturbation tend to be greater than those for the solar perturbation. The lunar current system finds its origin in the ionosphere dynamo and lies close to the dynamo height along with $S_q$ current system.

The dynamo electric field associated with the wind drives a current, which tends to converge in some regions of space and cause an accumulation of positive charge, while in other regions of space it would diverge and cause negative charge to accumulate. These charges would create an electric field, which would cause current to flow tending to drain the charges. An equilibrium state would be attained when the electric-field-driven current drained charge at precisely the rate it was being accumulated by the wind-driven current. A net current flows in the ionosphere owing to the combined action of the wind and electric field (Takeda and Maeda, 1980). A large-scale vortex current at middle and low-latitudes flows counter clockwise in the northern hemisphere and clockwise vortex flows in the southern hemisphere. Traditionally these vortices are known as the $S_q$ current system because of the nature of the ground-level magnetic field variations that they produce. Currents and electric fields produced by the ionospheric wind dynamo are relatively weak in comparison with those of the solar wind/magnetospheric dynamo at high latitudes. Electric field in the equatorial lower ionosphere has a localized strong enhancement of the vertical component associated with the strong anisotropy of the conductivity in the dynamo region. This enhanced electric field drives an eastward daytime current along the magnetic equator called equatorial electrojet (Forbes, 1981;

Richmond, 1986). Efforts are being made to understand the changes in equatorial electrojet in response to the electrodynamic processes involved in the coupling between the solar wind, magnetosphere and ionosphere. This is due to dynamo region electric fields being communicated to higher latitudes along the geomagnetic field lines. Monitoring the upper atmosphere by coherent and incoherent backscatter radar observations has confirmed that the distributions in the dynamo region electric fields at equatorial latitudes originate in the corresponding electrodynamics disturbances at high latitudes (Somayajulu et al., 1985). Studies based on the surface magnetic data have shown consistent and near instantaneous response of equatorial electrojet variations to geomagnetic disturbances at high latitudes (Rastogi and Patel, 1975). Ionospheric dynamo is also affected by the absorption of ozone at the lower altitudes (30-60 km) and presence of stronger winds at higher altitudes (> 130 kms).

Dynamo potential differences may increase during the geomagnetic storms period due to the enhanced E and F region winds. Geomagnetic disturbed ionospheric wind dynamo can produce potential differences comparable to those produced by the quiet time dynamo (~ 10 kV), with higher potentials at the equator than at high latitudes (Blanc and Richmond, 1979).

## 2.3 Magnetospheric dynamo

The stresses applied to the outer magnetosphere by the solar wind and dynamical processes in the tail, is ultimately applied to the terrestrial ionosphere and upper atmosphere. As a consequence of the high thermal energy of the plasma the drift motion causes charge separation and hence polarization electric field is setup directed from dawn to dusk. The currents flow along the geomagnetic field lines down into the ionosphere on the dawn side and up from the ionosphere on the dusk side, both foot points being electrically connected via the dynamo region. This process can be considered similar to a huge hydromagnetic generator situated in to the magnetosphere (in which kinetic energy of the solar wind plasma is converted in to electric energy) and the load in the ionosphere; linked to each other via field-aligned currents (Strangeway and Raeder, 2001). The work done by these currents in the ionosphere overcomes the drag on the flow over the polar cap away from the sun and on the flow back towards the sun at lower latitudes. The force on the plasma moving anti-sunward across the polar cap is supplied by the solar wind, predominantly by reconnection with the magneto-sheath magnetic field (Russel and Fleishman, 2002).

Mass, momentum and energy transfer from the solar wind via the magneto-sheath can occur through the cusps as a result of a number of processes, of which magnetic reconnection is the most important. Lester and Cowley (2000) have discussed the role played by reconnection in the magnetospheric convection and its importance for space weather. There are two processes by which the solar wind plasma can cross the magnetopause, (i) direct entry due to flow along reconnected open field lines (Dungey, 1961; Gonzalez et al., 1994) and (ii) cross field transport due to scattering across closed magnetopause field lines (Lee et al., 1994). The first process is more likely to be important when the interplanetary magnetic field (IMF) is directed southward. In this case, the solar wind and magnetospheric field lines are anti-parallel; the magnetic reconnection can occur easily leading to 5 to 10% solar wind energy imported into the earth's magnetosphere (Weiss et al., 1992) during sub-storms and storms. During northward IMF intervals, the energy injection due to magnetic reconnection is considerably reduced and cross-field transport becomes important. Tsurutani and Gonzalez (1995) have estimated that about 0.1 to 0.3% of the solar wind energy gets transferred to the magnetosphere during northward IMF. Several other processes, like impulsive penetration of the magneto-sheath plasma elements with an excess momentum density (Owon and Cowley, 1991), plasma entry due to solar wind irregularities (Schindler, 1979), the Kelvin Helmholtz instability (Miura, 1987) and plasma percolation due to overlapping of a large number of tearing islands at the magnetopause (Galeev et al., 1986) have been suggested for the plasma transport across the magnetopause.

The plasma-sheet, central part of the geo-magneto-tail is a giant plasma reservoir where the plasma (ionospheric / solar wind origin) is gathered and accelerated from a few eV to a few keV. The finite dawn-dusk electric field imposed by the solar wind magnetosphere interaction in the whole cavity accelerates the ions and to a lesser extent the electrons in the central part of the plasma sheet, where the magnetic field almost vanishes. The plasma in the plasma sheet has very large ratio $\beta$ between the kinetic and the magnetic pressure, this makes the system extremely unstable; very fast developing time variations called sub-storms, do develop in the plasma sheet. Prior to a sub-storm, the plasma sheet extends to a large space where the vertical component of the magnetic field becomes significant. The plasma is pushed earthward by $\mathbf{E} \times \mathbf{B}$ convection (**E** being essentially dawn to dusk) and the tail current increases. As plasma is convected inward, it faces increased magnetic field. The conservation of the adiabatic invariants leads to the energization of plasma by Fermi and betatron mechanism. The small-scale fluctuations that develop within the magnetospheric boundaries/discontinuities also play a significant role in accelerating electrons along field lines.

Energetic particles, especially electrons precipitate from the magnetosphere due to wave-particle interaction in to the upper atmosphere and produce the visible emissions called the aurora borealis (in the north) or the aurora austral is (in the south). This energetic particle precipitation also causes significant ionization, heating and dissociation in the thermosphere. Further, the energized plasma also has an important influence on the flow of the electric currents and on the distribution of electric fields *(*Spiro and Wolf, 1984*)*. Energetic particles drift in the Earth's magnetic field, electrons towards the east and positive ions towards the west, so that a westward ring current flows which exerts an electromagnetic force on the plasma directed away from the Earth; thus tending to oppose the earthward convection. Charge separation associated with the ring current tends to create an eastward electric field component, opposite to the night side westward convection electric field, largely canceling the convection electric field in the inner magnetosphere. The overall pattern of the magnetospheric convection tends to map along the magnetic field line in to the ionosphere even though this mapping is imperfect because of net electric field that tends to develop within the non-uniform energetic plasma. In the upper ionosphere,

$\mathbf{E}_{Plasma} = 0 = \mathbf{E} + \mathbf{v_s} \times \mathbf{B}$ (1)

where E is the electric field in the Earth fixed frame of reference, $v_s$ is velocity of solar wind plasma with respect to the Earth and B is geomagnetic field vector. The electric field strength in the auroral oval tends to be somewhat larger than the polar cap electric field. Dramatic disturbances of the entire magnetosphere during the magnetic storms period, lasting for about a day is produced by the enhancement of the solar wind velocity, density and southward IMF component (Roble, 1985; Richmond, 1986; Roble and Tzur, 1986). These storms are predominantly phenomena of the solar wind/magnetosphere dynamo, but they are also affecting the ionospheric wind dynamo. Auroral conductivity enhances due to the storm and these conductivities can also be altered at lower latitudes at night by over an order of magnitude (Rowe and Mathews, 1973). The entire wind system during the major storms in the dynamo region can be altered by the energy input to the upper atmosphere; the pattern of electric field generation is modified (Blanc and Richmond, 1979). The above discussion of the magnetospheric convection is a simplified picture and an indicative of the role played by it in the electromagnetic coupling of the upper atmosphere. The complete description is not possible in this review

## 3. Global Electric Circuit models

A few mathematical models of global atmospheric electric circuit involving various generators as discussed above have appeared over the years (Kasemir, 1977; Hill, 1971; Hays and Roble, 1979; Ogawa, 1985; Roble, 1991). Hays and Roble (1979) presented a quasi-static model, which is shown in Figure 1 that couples many of the elements operating in the global circuit. They considered thunderstorms as positive and negative point pairs constituting current sources that can be randomly distributed in various thunderstorm regions around the earth, including the effects of earth's orography and electrical coupling along geomagnetic field lines in the ionosphere and magnetosphere. However, they did not consider latitudinal, longitudinal and height variations of the atmospheric conductivity. Makino and Ogawa (1984) considered a numerical model including the conductivity details, but the distribution of aerosol particle concentration near the earth-surface and its subsequent effect on the global resistance is missing.

Sapkota and Varshneya (1990) studied the effects of pollution due to aerosol particle (anthropogenic and volcanic eruption), ionization caused by the coronae discharges, solar activity and of stratospheric aerosol particles (SAP) on the parameters of the GEC. They have shown that an increase in SAP increases the global resistance, while both global current and ionospheric potential decrease. The SAP affects the electrical structure of the stratosphere and the troposphere except in volcanically active region, where conductivity is low due to high aerosol particle concentration. A 7 % increase in the ionospheric potential by global variation of ionization due to solar activity has little effect on the ground electrical properties, where more than 30 % variations have been reported to be caused by local effects (Sapkota and Varshneya, 1990). The calculations are based on the assumption that man made pollution has been increased substantially due to the activity in the northern hemisphere.

The widely referred model of Ogawa (1985), considering the simple equivalent circuit for the atmosphere and an equipotential surface for the ionosphere treats the thundercloud as a constant current generator with a positive charge at the top and negative charge at its bottom. Rycroft et al. (2000) presented a new model of GEC treating the ionosphere and the magnetosphere as passive elements and presented three different regions of fair-weather circuit. One of these is for the high-altitude part of the earth, where the profiles of J and E through the fair-weather atmosphere will differ from those of low and mid-latitudes. The energy associated with the global electric circuit is enormous $\approx 2 \times 10^{10}$ J (Rycroft et al., 2000). This value is obtained by considering 200 C charge associated with each storm and 1,000 storm operating around the globe at a time. The electric current density through the fair-weather atmosphere is taken as $\approx 2 \times 10^{-12}$ A/m$^2$. Taking conductivity of air at ground level to be $\approx 2 \times 10^{-14}$ mho/m, the fair weather electric field is $\approx 10^2$ V/m at the ground level, $\approx 1$ V/m at 20 km altitude, and $\approx 10^{-2}$ V/m at 50 km altitude (Rycroft et al., 2000). Thus, even though the fair-weather current remains the same, the vertical electric field goes on decreasing with altitude because of change in conductivity. For example, following a Forbush decrease, if the atmospheric conductivity is everywhere reduced by 10 %, then fair-weather electric field will be increased by ~10 % (Ogawa, 1985). While discussing the effect of sprites, Rycroft et al. (2000) argued that the ionospheric potential would reduce to 99 % of the initial value only for few milliseconds after sprites and would have little effect on the fair-weather electric field.

Harrison (2005) studied the average properties of the GEC. Rycroft (2006) has updated his GEC model, in which he included some new generators (i.e. mesospheric generators), along with some switches, which are closed for short time (when certain types of discharges occur).

In all GEC models, electrostatic phenomena have been considered, whereas during lightning discharges, electromagnetic waves having frequencies from a few Hz to 100 MHz are generated and propagated through the atmosphere. To account for the effect of these waves, electrodynamics/ electromagnetic effects should be considered by relating electromagnetic fields to charge and current densities in a time varying situation. At higher frequencies ($\omega \gg \sigma/\varepsilon_0$, where $\sigma$ is conductivity of the medium and $\varepsilon_0$ is permittivity of the free space), the medium can be considered as a leaky dielectric, whereas at lower frequencies ($\omega \ll \sigma/\varepsilon_0$), it can be considered as a conductor. Even in the absence of radiation, displacement current should be considered. In fact the Maxwell current (Singh et al., 2004) is very variable, and not a great deal is known about it. Further, optical emissions are associated with transient currents/ fields, which change the electrical properties of the medium. All these points should be included in the GEC model. The conductivity and columnar resistance of the ambient medium along with variations in the source, govern the electric field and small variation in it.

## 4. Electrical conductivity and columnar resistance of the GEC

The electrical conductivity depends upon the distribution of ions, electrons and presence of magnetic field in the ambient medium. Principle source of ionization in the lower atmosphere is the galactic cosmic rays, which maintains the electrical conductivity from the ground to about 60 km in altitude. Near the ground some additional ionization is produced due to release of radioactive gases from the soil and above 60 km solar ultraviolet radiation becomes important source. However, during the magnetic storm period ionization due to energetic auroral electron precipitation and auroral X-rays, Bremsstrahlung radiation along with proton bombardment (during solar proton events), also become significant sources specially in the high-latitude atmosphere.

The electrical conductivity increases roughly exponentially with altitude with a scale height ~ 7 to 8 km in the lower atmosphere due to the increase of cosmic ray's energy spectrum with altitude (Roble and Tzur, 1986) and charged particles precipitating from the magnetosphere. In the stratosphere, conductivity scale height is ~7 km. Hu (1994) showed that average positive conductivity was 15 % higher than the negative conductivity and it is the function of latitude but not longitude. The average vertical profile of the conductivity at the south pole has a scale height of ~ 10 km (Holzworth, 1991). When this result is compared with the ~ 7 km value obtained at other Antarctic and southern locations, it appears that the conductivity scale height may increase with increasing geomagnetic latitude across the polar cap.

The atmospheric conductivity increases form $10^{-13}$ to $10^{-7}$ mho/m when measured from the earth's surface to 80 km altitude (Cho and Rycroft, 1998). Hale (1994) presented a more complex profile depicting variation in both space and time under different geophysical conditions. Pasko and George (2002) discussed nighttime distribution of middle atmospheric conductivity both low for latitude and mid latitude conditions, which are very similar to that of Hale (1994). Radioactivity of the ground and its emanations cause significant variations in electrical conductivity near the ground both in space and time in an unpredictable way (Hoppel et al., 1986, Volland, 1987). One consequence of the increase in conductivity with altitude is that the columnar resistance of the atmosphere is concentrated near the surface. Vertical current flows between the positively charged ionosphere and the earth's surface. The current is closely linked to the vertical variations of aerosol and ion concentrations in the atmosphere, which together determine the total electrical resistance of the atmosphere.

The variation of columnar resistance at different latitudes due to geomagnetic influence on cosmic ray ion production and spatial changes in tropospheric aerosol modifying the ion removal rate, has been discussed by Harrison (2005). Global variation of columnar resistance is not well known, although it is an important property of the global electric circuit. In the highly polluted area the columnar resistance increases significantly and lowers the air-earth current. The columnar resistance integrated over the earth's surface provides the load term in global atmospheric electric circuit, whereas the integrated value over altitude provides a local value, which heavily depends on atmospheric, ion-aerosol interaction (Harrison and Carslaw, 2003). The phenomena involved in controlling the local value of columnar resistance is important as the physics of cosmic rays, ions, aerosols and clouds have been suggested to provide a mechanism linking solar change and climate (Carslaw et al., 2002).

At the top of the middle atmosphere the conductivity becomes anisotropic with the Pedersen conductivity ($\sigma_P$, carried by electrons below 100 km and by ions above that altitude) parallel to E-field and orthogonal to $B_0$, the Hall conductivity ($\sigma_H$, mainly due to electrons), orthogonal to E and $B_o$, and the field-aligned conductivity $\sigma_F$ parallel to $B_0$, because of the influence of the geomagnetic field and shows diurnal variation due to solar photo-ionization process. Pedersen and Hall conductivity peaks in the height range between 100 and 150 km, the dynamo region. As a result of large field aligned conductivity, the geomagnetic field lines behave like electric equipotential lines and hence electric field parallel to $B_0$ breaks down within a fraction of a second. Significant current flows if electric fields orthogonal to $B_0$ exist and Pedersen and Hall conductivities are large. The finite conductivity and its variation in space and time modify the transmission characteristics of the electromagnetic energy which is necessary for the interpretation of observed wave forms with respect to their original wave structure at the source (Volland, 1987).

Solar activity influences the conductivity on the day-to-day basis and decadal time scale with relative amplitude of 3–20 %. With an increase in solar activity, the GCR flux reduces in mid-latitude causing reduction in conductivity in this region, while during the same period solar protons may be 'funneled' by the Earth's magnetic field to the polar regions resulting in an increased conductivity there. The interaction of solar wind with the Earth's magnetic field also causes a dawn-to-dusk potential difference across the polar region (Tinsley and Heelis, 1993). During the active geomagnetic periods, the energetic charged particles precipitating from the inner and the outer Earth's magnetospheric radiation belts interact with the middle and the lower atmosphere by depositing their energy in the atmosphere and producing ionization directly or via Bremsstrahlung radiation, thereby influencing the dynamics of storm and atmosphere (Tinsley and Heelis, 1993; Tinsley, 2000*)*.

Markson and Muir (1980) suggested how solar variability moderates the Earth's electric field and electrical potential of the ionosphere, which is maintained by the world-wide thunderstorm activity. It also affects the weather and climate (Markson,1978), thus leading to a connection between electrical properties of the medium and weather and climate. Such a link supports the mechanism in which solar control of ionizing radiation modulates atmospheric electrification, cloud physical processes and atmospheric energy budgets. On the other hand, some tropospheric disturbances are known to influence the ionospheric phenomena. For example, several theoretical and experimental studies show that the lightning activity in thunderstorms influence the temperature, ion densities, composition and electrical potential of the ionosphere (Inan et al., 1991; Taranenko et al., 1993; Pasko et al., 1997). In Figure 4, it is shown that the solar activity along with tropical thunderstorms, control the ionosphere-earth current density, which is an important parameter in global atmospheric electric circuit. GEC can provide a good framework for understanding the solar-terrestrial weather relation. It relates the solar sector boundary crossing to the increasing lightning frequency (Reiter, 1972), thunderstorm activity (Cobb, 1967; Reiter, 1972) and vorticity area index (Markson, 1978). Lightning frequency and the tropospheric electric field are found to increase shortly after solar flares. On a longer time scale, highly positive correlation between the 11-yr sunspot cycle and thunderstorm activity has been reported (Stingfellow, 1974). Schlegel et al. (2001) extended this work globally and studied how a solar activity signal can be transmitted to lower atmosphere and argued that planetary waves may play a crucial role in it. Large horizontal potential drops in the ionospheric correlate well with the solar flare occurrences with the dealy of about 2 days or less (Muhleisen, 1977). Markson (1978) suggested that the atmospheric electrical response to solar activity may provide important clues to how the sun variations modulate weather. The current in non-thunderstorm cloud causes space-charge generation by transferring charge to aerosol particles and droplets. Evaporation of droplets concentrates charges and leads to electro-scavenging (Tinsely, 2000; Tripathi, 2000, Tripathi and Harrison, 2001,2002). The scavenging of aerosol particles may lead to changes in the concentration of condensation nuclei which can cause changes in the indirect aerosol affect of cloud cover and precipitation rates. Such changes can have weather and climate consequences (Carslaw et al., 2002; Kniveton and Tinsley, 2004, Tinsley et al., (2006) have discussed electrically enhanced scavenging, and the electrical inhibition of scavenging in the context of the microphysics of weakly electrified clouds.

**5. Global Electric Circuit and global temperature (climate change)**

Recently it has been suggested to use GEC as a tool for studying the earth's climate and climatic changes, because of its direct connection with lightning activity (Williams, 2005). Williams (1992) reported an extremely non-linear increase in tropical lightning rate when temperature rose above critical threshold (nonlinear sensitivity of thunderstorm activity to temperature). He also showed high correlation between monthly mean of tropical surface air temperature and SR measurements of global lightning activity. Fullekrug and Fraser-Smith (1998) have inferred global lightning and climate variability from the ELF magnetic field variations. In the global frame work the response of lightning and electrified clouds to temperature and change in temperature have been analyzed and many time scales including semi-annual, annual, etc have been reported (Williams, 2005, and references there in). In the case of semi-annual variation, even $1^0C$ increase in temperature may result in 50% increase in global lightning frequency (Williams, 2005).

Price and Rind (1992) parameterized global lightning activity using satellite cloud data from the ISCCP (International Satellite Cloud Climatology Project) and predicted more lightning in a warmer world due to enhanced $CO_2$ content. Markson and Price (1999) reported positive correlation between ionospheric potential and global temperature, whereas ionospheric potential was positively correlated with an inferred global lightning/deep cloud index, which is also positively correlated with global temperature. They suggested that warmer temperatures lead to more deep convection resulting in higher ionospheric potential. Price (1993) showed good agreement between the diurnal surface temperature changes and the diurnal variability of GEC. He suggested that a 1 % increase in global surface temperature could result in a 20 % increase in ionospheric potential. Therefore, the above study showed a strong link between the frequency/intensity of global deep convection and global surface temperatures. Muhleisen (1977) and Markson (1986) suggested that measured ionospheric potential agree well with the "Carnegie Curve" confirming that lightning plays a major role in the global electric circuit. Both ac and dc global circuits have been found to respond to global temperature changes on time scales ranging from the diurnal (Price, 1993), through seasonal (Price, 1994; Williams, 1994) to the El Nino-Southern Oscillation scales (Williams, 1992).

The observed signal level in SR suggests that the cloud to ground lightning is not the sole driver of the global electric circuit (Williams, 1992; Williams and Heckman, 1993). Markson and Lane-Smith (1994) suggested that combination of SR power level monitoring and regular ionospheric soundings could be used to infer proxy measures of both global temperature and global rainfall rates. A close relationship has been shown between: (i) tropical surface temperature and monthly variability of SR (Williams, 1992; 1994); (ii) ELF observations in Antarctic/Greenland and global surface temperature (Fullekrug and Fraser-Smith, 1998); (iii) diurnal surface temperature changes and the diurnal variability of the GEC (Price, 1993); and (iv) ionospheric potential and global/tropical surface temperature (Mulheisen, 1977; Markson, 1986; Markson and Price, 1999). It has led to the speculation that global warming would result in enhanced convective activity, which may result in increased thunderstorm production on a global scale. Reeve and Toumi (1999) using satellite data, showed agreement between global temperature and global lightning activity. Price (2000) extended this study and showed a close link between the variability of upper troposphere water vapor (UTWV) and the variability of global lightning activity. Also UTWV has excellent agreement with surface temperature and lightning activity through the measurement of the SR. This is an important finding in the area of lightning and climate. UTWV is closely linked to the other phenomena such as tropical cirrus cloud, stratospheric water vapor content, and tropospheric chemistry (Price, 2000). These examples suggested that by monitoring the GEC, it is possible to study the variability of surface temperature, tropical deep convection, rainfall, upper troposphere water vapor content, and other important parameters, which affect the global climate system.

The global warming issue has dominated the area of climate research for many years. In this connection, large developments have appeared with new data sources for global lightning, both optical and radio frequency. Many time scales have been explored for lightning variations such as diurnal, inter-seasonal, semiannual, annual and inter-annual, dominated by the ENSO (Williams, 2005). Williams (1992) found a doubling in SR amplitude over the 1992 ENSO (El Nino Southern Oscillation) event that correlated with the tropical temperature anomaly (Hansen and Lebedeff, 1987). Systematic changes in the meridional location of tropical thunderstorm regions was inferred from the observations of SR frequency variation on the ENSO time scale (Satori and Zieger, 1999), a positive inter-annual correlation but the lightning changes were extra tropical as reported by Reeve and Toumi (1999). Recently, Williams et al. (2005) have discussed in detail the physical mechanisms and hypotheses linking temperature and thermodynamics with lightning and global circuit. This clearly shows that the study of physical processes involved in the global electric circuit, the variability of global lightning activity and its relation to surface temperatures, tropical deep convection, rainfall, upper tropospheric water vapor content, and other important parameters that affect the global circuit are essential to understand the dynamics of biosphere, which is essential for the betterment of the human society.

## 6. Global Electric Circuit and Aerosols

The electrical conductivity in clean atmosphere is inversely proportional to aerosol particle content in the air. The conductivity is therefore considered as an index of atmospheric aerosol loading over the open ocean and has been used to estimate global changes in the background air pollution level (Cobb and Wells, 1970; Retalis and Retalis, 1998; Kamra et al., 2001). Harrison and Aplin (2002, 2003) used a theoretical model for the development of convective boundary layer to determine the domination of local pollution effect on the atmospheric electrical potential measurements. Using this technique and atmospheric electrical data along with the theory of boundary layer meteorology, pollution concentration and its composition can be derived. Harrison and Aplin (2002) estimated surface pollution concentration to be 60±30 µg m$^{-3}$ at Kew in 1863. This value is substantially lower than the previously derived value. They concluded that the diurnal variations in smoke pollution differ between the seasons, and have changed their character after the advent of motor traffic.

Aerosols in the atmospheric boundary layer and stratosphere have a strong influence on the electrical phenomena in the atmosphere. Adlerman and Williams (1996) found large effect from several factors such as seasonal changes, variations in mixed layer heights, variations in the production rates and anthropogenic aerosols and variation in surface wind speed on the seasonal variations of GEC. The high concentration of aerosols decrease the conductivity of the air in the boundary layer, and affect the electrical structure of the lower atmosphere (Manes, 1977; Morita and Ishikawa, 1977). Sapkota and Varshneya (1990) have studied the effect of aerosol particles of anthropogenic and volcanic origins on the global electric circuit. It reduces the current density substantially. They have developed a model for the distribution of aerosol particles based on distribution of world population density and computed columnar resistance, current density, potential distribution and electric field.

The impact of the Chernobyl nuclear power plant accidents in April 1986 on all atmospheric electrical parameters in some parts of Sweden was observed to be significant even several months later (Israelsson and Knudsen, 1986). Martell (1985) and Israelsson et al. (1987) have studied the effect of radioactive material injected into the atmosphere on the production of lightning flashes. The increased radiation enhances the ionization in the atmosphere, which affect the charging mechanism of clouds leading to enhanced thunderstorm activity as was observed after Chernobyl accidents. Israelsson et al. (1987) have proposed that enhanced ion production rate increases the air conductivity surrounding the cloud, which may result in to rapid development of screening layer if charge separation process exists inside the cloud. This may lead to rapid reduction of the electric field external to the cloud. On the other hand if the air conductivity is appreciably increased by the radioactivity then the electric field inside the cloud i.e between the cloud charges and the screening layer charges will be enhanced. This may lead to an increased lightning activity. In the light of precise measurements now available, the above conjecture needs to be examined.

In recent years, innovative methods have been developed to diagnose and study cloud microphysics specially in quantifying the role of aerosol as mediator of cloud microphysics, precipitation, cloud electrification and lightning (Rosenfeld and Lensky, 1998; Rosenfeld and Woodley, 2003). An increase in aerosol concentration may lead to a reduction in mean droplet size, a suppression of warm rain coalescence and an enhancement of the cloud water reaching the mixed phase region (Williams et al., 2000). Steiger and Orville (2003) reported lightning enhancements over oil refineries near Lake Charles and Louisians and interpreted it as an aerosol effect. However, the role of surface properties in controlling the lightning activity was considered to explain the sharp discontinuity of flash density at the Texas and Louisians coastlines. Measurements of Andreae et al. (2004) suggested that aerosols enhance the convection process. Jungwirth et al. (2005) have suggested that the aerosol chemistry should also be included in explaining the cloud-to-ground lightning activity. This shows that the distribution of shape, size and concentration of aerosols in a complex way affect the lightning activity and require systematic studies in relation to GEC and space-weather/climate.

## 7. Global Electric Circuit and Cosmic Rays

Cosmic rays ionize the atmospheric gas constituents and hence modify the atmosphere's columnar resistance and ionospheric potential (Markson, 1981). The effect of cosmic rays on ionospheric potential was originally identified from solar modulation of lower energy cosmic rays in the form of eleven-year cycle. The correlation between galactic cosmic rays and cloud cover an 11-year solar cycle basis (Carslaw et al., 2002) has enhanced the interest in the study of relationships between solar variations and weather changes. The cosmic ray flux penetrating down to the earth's lower atmosphere during active period of the Sun decrease due to its interaction with solar wind. The steady increase in solar activity during the twentieth century has led to a secular decline in cosmic rays (Carslaw et al., 2002) and the expected global circuit response has been identified in surface measurements of potential gradient in the United Kingdom (Märcz and Harrison, 2003), although aerosol changes have also been suggested to be responsible agent (Williams, 2005). Harrison and Ingram (2005) have also reported a decrease in the air- earth current at Kew and in the potential gradient in mountain air in the 1970s.

The cosmic rays may also affect the climate / weather involving cloud processes such as condensation of nucleus abundances (Wilcox et al., 1974), thunderstorms electrification and thermodynamics (Markson and Muir, 1980), ice formation in cyclones (Tinsley, 1996) etc. Svensmark and Friis-Christiansen (1997) have analyzed data and showed a correlation between cosmic rays and earth's cloud cover over a cycle. Turco et al. (1998) and Marsh and Sevensmark (2000) suggested that galactic cosmic rays could generate aerosol particles that can act as cloud condensation nuclei and affect, particularly over ocean, the formation and thickness of cloud. They also found a strong association between low clouds, at around 3 km altitude, and cosmic rays flux. Thus, it is likely that the cosmic ray influences the GEC as well as climate/ weather. This early suggestion of Ney (1959) still remains to be explored in depth. In fact the observation of a correlation between cosmic ray intensity and cloudiness provides an opportunity to investigate ion-aerosol-cloud interaction, because variation in ion production rate due to cosmic rays may impact aerosol distribution and cloud formation. In fact rainfall is an important controlling factor of average cloudiness of a region via cloud lifetime. Even ice-particles growth induces rainfall because liquid clouds are highly supersaturated with respect to ice. Both these aspects can be studied in detail through ion-aerosol-cloud interactions in which ion production is governed by galactic cosmic rays.

## 8. Importance of the GEC study to our society

Global electric circuit and global lightning activity and its relation with the global climate form a basis of the proposal for analyzing/ studying the issues involved in climate studies and variations in it. Global signals are evident on many different time scales and forcing mechanism for many of these time scales are known. For example, in discussing the coupling between global electric circuit and general circulation of the atmosphere, it is essential to distinguish between latent heating/ rainfall and electrification/ lightning because the former is prevalent with shallow, gentle lifting of air, whereas the latter is caused by deeper and stronger lifting. Thus, detailed study of atmospheric circulation is essentially required both for understanding GEC and climate.

GEC is controlled by the space-weather parameters such as solar wind, solar flares, coronal mass ejections etc. through geomagnetic activity. As a consequence of enhanced solar activity, current surges can be induced in power lines, causing flickering lights and blackouts resulting in huge damage. Further, telecommunication cables and petroleum pipelines are also affected. The serious consequence of bad space weather is disruption of satellite communication and satellite links. Even, there is possibility of damage to earth-orbiting satellites. Here, it should be noted that these effects are not mediated by the global electric circuit, although, they are the consequences of solar variations. Further studies are required in this direction, because solar influence (on various time scale) on temperature, thunderstorms frequency, tropopause heights, atmospheric circulation, occurrence of drought etc is known (Carslaw et al., 2002 and references there in). Solar variability affect the earth's weather in the following way namely enhanced solar irradiance may provide excess heat input to the lower atmosphere leading to global warming, solar ultraviolet may be absorbed in the lower atmosphere and change the local electrodynamics of the region. The other root is through the cosmic rays, which control both weather as well as global electric circuit.

In general, the study of various problems related to global electric circuit is definitely beneficial to our society because (i) the study explores inter-connection between the electrical environment, climate and weather of the earth's atmosphere. It may be possible in the future to modify climate and weather by controlling some of the electrical parameters of the global electrical circuit (Bering III, 1995). This point has been briefly discussed in section 5. (ii) The detail study of global electric circuit and its response to extreme solar and geomagnetic variability is useful in modeling of biosphere. The role of the global electrical circuit under extreme conditions of solar variability may be used in future to assess the harmful effect of the latter. (iii) The close association of atmospheric circulation and GEC makes one to speculate / explore link between GEC and global warming by $CO_2$. However, there is no observational evidence for such a relation to exist. (iv) The aesthetic requirement of understanding the electrical behavior of the Earth's environment (Singh et al., 2004) in the same ways as one would like to understand meteorological parameters; (v) The detailed knowledge of the forcing from the global electric circuit would be useful in evaluating the electrical response of the planetary boundary layers (PBL), because electrical processes in this region are complex, highly variable and span a tremendous range of space and time scales. The high variability of aerosol particle concentration, size composition leads to one of the largest uncertainties of anthropogenic climate forcing. Even the downward movement of planetary boundary layer causes enhanced pollution in the region. This is evident in the winter season in polluted cities. Since the dynamics of planetary boundary layer is affected by GEC, the detailed study of GEC finds importance in human welfare. (vi) Better estimates of global lighting activity will help in more accurate estimates of the production rates of $NO_x$ on global scale, a vital factor for understanding of the climatic changes and ozone hole, and hence in global warming.

9. **Recommended areas for future work**

Based on the discussions presented above, the following problems are recommended for further investigations:

(i) The thunderstorms and galactic cosmic rays mainly control the global electric circuit and link it with the climate and variations in it on different time scales. Discussion in Section 5 brings out how this linkage is reflected in various atmospheric manifestations such as convection, lightning, global surface temperature, SR frequency, UT WV etc. Therefore, it becomes imperative to investigate the temporal variations of global electric circuit and the mechanisms responsible for such variations on different time scales. Further, keeping in view that a single measurement of SR has the potential of replacing the measurements of temperature all over the globe, it is important to investigate the possibility that whether GEC and the global/regional lightning frequency can act as an indicator of climate change or not.

(ii) To study the nature and sources of middle-atmosphere discharges to (a) increase knowledge of the characteristics of the recently discovered phenomena such as sprites, blue jet, elves etc and their possible association with severe weather, (b) to explore their effects on radio wave propagation and atmospheric chemistry, (c) to quantify global occurrence rate of sprites and the modification of ionospheric potential by them, and (d) to quantify the current to be injected in to GEC by the optical phenomena.

(iii) Some of the non-linear processes associated with cloud electrification mechanism are not clear. An attempt should be made to elucidate the fundamental physics of these processes involved in lightning.

(iv) The explanation of observed optical phenomena such as sprite, elves etc require production of transient current/electric fields in the mesosphere, which require further studies.

(v) To quantify the production of oxides of nitrogen ($NO_X$) by lightning to better understand the upper-tropospheric production or loss of ozone, for better estimation of the global warming.

(vi) To investigate the mechanism, if any, by which cosmic rays affect clouds and hence the weather and climate.

(vii) To quantify the significant feedback processes at work in the electrically coupled atmosphere- ionosphere-magnetosphere system.

(viii) The role of global warming (increasing green house effect) on the various phenomena/ processes involved in global atmospheric electric circuit and its coupling to climate.

**Acknowledgement**

The authors thank the reviewer for constructive suggestions. The work is also partly supported by Department of Science & Technology (DST) Government of India, under SERC project. (DS) thanks to DST, under the BOYSCAST programme (with reference SR/BY/A-19/05) and also Dr Urmas Horrak for the full support.

## References

Adlerman, E. J., Williams, E. R., 1996. Seasonal variation of the global electric circuit. J. Geophys. Res. 101, 29679-29688.

Andreae, M.O., Rosenfeld, D., Artaxo, P., Costa, A.A., Frank, G.P., Longo, K.M., Silva-Dias, M.A.F., 2004. Smoking rainclouds over the Amazon. Science 303, 1337-1342.

Barr, R, Llanwyn, J.D., Rodger, C.J., 2000. ELF and VLF radio waves. J. Atmos. Solar-Terr. Phys. 62, 1689-1718.

Barrington-Leigh, C.P., Inan, U.S., Stanley, M., Cummer, S.A., 1999. Sprites triggered by negative lightning discharges. Geophys. Res. Lett. 26, 3605-3608.

Bering III, E.A., 1995. The global circuit: Global thermometer, weather by product or climate modulator? Rev. Geophys. (supplement copy Part II) 845-862.

Bering III, E.A., Few, A. A., Benbrook, J.R., 1998. The global electric circuit. Phys. Today (Oct. issue), 24-30.

Bering III, E.A., Benbrook, J.R., Garret, J.A., Paredes, A.M., Wescott, E.M., Moudry, D.R., Sentman, D.D., Stenback-nielsen, H. C., 2002. The electrodynamics of sprites. Geophys. Res. Lett. 29, 10.1029/2001/GL 013267.

Blanc, N., Richmond, A. D., 1979. The ionosphere distribution dynamic. J. Geophys. Res. 85, 1669-1686.

Boccippio, D.J., Williams, E.R., Heckman, S.J., Lyons, W.A., Baker, I.T., Boldi, R., 1995. Sprites, ELF transients and positive ground strokes. Science 269, 1088-1091.

Carslaw, K.S., Harrison, R.G., Kirkby, J., 2002. Cosmic rays, clouds, and climate. Science 298, 1732-1737.

Cho, M., Rycroft, M.J., 1998. Computer simulation of electric field structure and optical emission from cloud top to ionosphere. J. Atmos. Solar Terr. Phys. 60, 871-888.

Cobb, W. E., 1967. Evidence of a solar influence on the atmospheric electric element at Mauna Lao observatory. Mon. Weather Rev. 95, 905.

Cobb, W.E., Wells, H.J., 1970. The electrical conductivity of oceanic air and its correlation to global atmospheric pollution. J. Atmos. Sci. 27, 814-819.

Cummer, S. A., Inan, U. S., Bell, T. F., and Barrington-Leigh, C. P., 1998. ELF radiation produced by electrical currents in sprites. Geophys. Res. Lett. 25, 1281-1285.

Dungey, J. W., 1961. Interplanetary magnetic field and auroral zones. Phys. Rev. Lett., 6, 47-48.

Fishman, G., Bhat, P.N., Mallozzi, R., Horack, J.M., Koshut, T., Kouveliotou, C., Pendleton, G.N., Meegan, C.A., Wilson, R.B., Paciesas, W.S., Goodman, S.J., Christian, H.J., 1994. Discovery of intense gamma rays flashes of atmospheric origin. Science, 264, 1313-1316.

Forbes, J.M., 1981. The equatorial electrojet. Rev. Geophys. Space Phys. 19, 469-504.

Fullekrug, M., Fraser-Smith, A. C., 1998. Global lightning and climate variability inferred from ELF magnetic field variations. Geophys. Res. Lett. 25, 2411-2414.

Galeev, A. A., Kuznetsova, M. M. Zelenyi, L. M., 1986. Magnetopause stability threshold for patchy reconnection. Space Sci. Rev., 44, 1.

Gonzalez, W. D., Joselyn, J. A., Kamide, Y., Krochl, H. W., Rostoker, G., Tsurutani, B. T., Vasyliunas, V. M., 1994. What is geomagnetic storm? J. Geophys. Res. 99, 5771.

Hale, L.C., 1987. Terrestrial atmospheric electricity. In: Encyclopedia of Physical Science and Technology. Ist ed. Vol. 13, pp 736.
29

Hale, L.C., 1994. The coupling of ELF/VLF energy from lightning and MeV particle to the middle atmosphere, ionosphere and global circuit. J. Geophy. Res. 99, 21089-21096.

Hansen, J.E., Lebedeff, S., 1987. Global trends of measured surface air temperature. J. Geophys. Res. 92, 13345-13372.

Harrison, R.G., 2005. The global atmospheric electric circuit and climate. Survey in Geophys. 25, 441-484.

Harrison, R.G., Alpin, K. L., 2002. Mid-nineteenth century smoke concentrations near London, Atmos. Environ. 36, 4037-4043.

Harrison, R.G., Alpin, K.L., 2003. Nineteenth century Parisian smoke variation inferred from Eiffel Tower atmospheric electrical observations, Atmos. Environ. 37, 5319-5324.

Harrison, R.G., Carslaw, K.S., 2003. Ion-aerosol-cloud processes in the lower atmosphere. Rev. Geophys. 41, 10.1029/2002RG000114.

Harrison, R.G., Ingram, W.J., 2005. Air-earth current measurements at Kew, London. Atmos. Res. 76, 49-64.

Hays, P. B., Roble, R. G., 1979. A quasi-static model of global atmospheric electricity 1 The lower atmosphere. J. Geophys. Res. 84, 3291-3305.

Heckman, S.J., Williams, E.R., Boldi, R., 1998. Total global lightning inferred from Schumann resonance measurements. J. Geophys. Res. 103, 31,775-31779.

Hill, R. D.: 1971. Spherical capcitor hypothesis of the Earth's electric field. Pure and Applied Geophys. 84, 67-75.

Hiraki, Y., Lizhu, T., Fukunishi, H., Nambu, K., Fujiwara, H., 2002. Development of a new numerical model for investigation the energetic sprites. Eos. Trans. AGU 83 (47), Fall Meet. Suppl. Abstract A11 C-0105.

Holzworth, R.H., 1991. Conductivity and electric field variations with altitude in the stratosphere. J. Geophys. Res. 96, 12,857-12,864.

Holzworth, R. H., Hu, H., 1995. Global electrodynamics from superpressure balloons. Adv. Space Res. 16, 131-140.

Hoppel, W.A., Anderson, R.V., Willett, J.C., 1986. Atmospheric electricity in the planetary boundary layer. In Study in Geophysics-The Earth electrical Environment National Academy Press, Washington, D.C, pp 149-165.

Horrak, U., 2001. Air ion mobility spectrum at a rural area. Ph.D thesis, University of Tartu, Tartu, Estonia.

Hu, H., 1994. Global and Local Electrical Phenomena in the Stratosphere. Ph.D. dissertation, Univ. of Washington, Washington.

Hu, H., Li., Q., Holzworth, R. H., 1989. Thunderstorm related variations in stratospheric conductivity measurements. J. Geophys. Res. 94, 16429-16435.

Huang, E., Williams, E., Boldi, R, Heckman, S., Lyons, W., Taylor, M., Nelson, T., Wong, C., 1999. Criteria for sprites and elves based on Schumann resonance observations. J. Geophys. Res. 104, 16943-16964.

Inan, U.S., Bell, T.F., Ridriguez, J.V., 1991. Heating and ionization of the lower ionosphere by lightning. Geophys. Res. Lett. 18, 705-708.

Israelsson, S., Knudsen, E., 1986. Effect of radioactive fallout from a nuclear power plant accident on electrical parameters. J. Geophys. Res. 91, 11909-11910.

Israelsson, S., Schutte, T., Pisler, E., Lundquist, S., 1987. Increased occurrence of lightning flashes in Sweden during 1986. J. Geophys. Res. 92, 10996-10998.

Jones, D.L., 1999. ELF sferics and lightning effects on the middle and upper atmosphere. Modern Radio Science 171-189.

Jungwirth, P., Rosenfeld, D., Buch, V., 2005. A possible molecular mechanism of thundercloud electrification. Atmos. Res. 76, 190-205.

Kamra, A.K., Murugavel, P., Pawar, S.D., Gopalakrishnan, V., 2001. Background aerosol concentration derived from the atmospheric conductivity measurements made over the Indian ocean during INDOX. J. Geophys. Res. 106, 28643-28651.

Kasemir, H. W., 1977. Theoretical problem of the global atmospheric electric current, Electrical Processes in Atmospheres. In Dolezalek, H., Reiter, R. (eds), Steinkopff Verlag, Darmstadt, pp 423-438.

Kniveton, D.R., Tinsley, B.A., 2004. Daily changes in global cloud cover and Earth transits of the heliospheric current sheet. J. Geophys. Res. 109, D11201,doi:10.1029/2003D004232.

Krider, E., Musser, J.A., 1982. The Maxwell current density under thunderstorm. J. Geophys. Res. 87, 11171-11176.

Lakhina, G.S., 1993. Electrodynamic coupling between different regions of the atmosphere. Current Science 64, 660-666.

Lee, L. C., Johnson, J. R. Ma, Z. W., 1994. Kinetic Alfven wave as a source of plasma transport at the dayside magnetopause. J. Geopgys. Res., 99, 17405.

Lester, M., Cowley, S. W. H., 2000. The solar terrestrial interaction and its importance for space-weather. Adv. Space Res. 26, 79-88.

Lyons, W., 1996, Sprites observations above the US high plains in relation to their parent thunderstorms systems. J. Geophys. Res. 101, 29641-29652.

Makino, M., Ogawa, T., 1984. Response of atmospheric electric field and air-earth current to variations of conductivity profiles. J. Atmos. Terr. Phys. 46, 431-445.

Manes, A., 1977. Particular air pollution trends deduced from atmospheric conductivity measurements at Bet-Dagan (Israel). In: Electrical Processes in Atmospheres. (Eds) Dolezalek, H., Reiter, R., Dr. Dietrich Steinkopff Verlag, Darmstadt, pp 109-118.

Marcz, F., Harrison, R.G., 2003. Long-term changes in atmospheric electrical parameters observed at Nagycenk (Hungary) and UK observatories at Eskdalemuir and Kew. Ann. Geophys. 21, 2193-2200.

Markson, R., 1978. Solar modulation of atmospheric electrification and possible implications for the Sun-weather relationship. Nature 273, 103-109.

Markson, R., 1981. Modulation of the earth's electric field by cosmic radiation. Nature 291, 304-308.

Markson, R., 1986. Tropical convection, ionospheric potential and global circuit variations. Nature 320, 588-594.

Markson, R., Muir, M., 1980. Solar wind control of the Earth's electric field. Science 206, 979-990.

Markson, R., Lane-Smith, D.R., 1994. Global change monitoring through the temporal variation of ionospheric potential. In preprint volume, Fifth Symposium on Global Change Studies and the Symposium on Global Electric Circuit, Global change and the meteorological Applications of Lightning Information, Ann. Meeting AMS, Nashville, TN, Jan. 1994, Am. Meteor. Soc., Boston, pp. 279-287.

Markson, R., Price, C., 1999. Ionospheric potential as a proxy index for global temperatures. Atmos. Res. 51, 309-314.
Marsh, N., Svensmark, H., 2000. Cosmic rays, cloud, and climate. Space Sci. Rev. 94, 215-230.
Martell, E.A., 1985. Enhanced ion production in convective storms by transpired radon isotopes and their decay products. J. Geophys. Res. 90, 5909-5916.
Matsushita, S., 1967. Solar quiet and lunar daily variation fields. In: Matsushita, S., Campbell, W.H. (Eds), Physics of Geomagnetic Phenomena. Academic Press, New Yark,, pp. 301.
Milikh, G., Valdivia, J.A., 1999. Model of gamma rays flashes fractal lightning. Geophys, Res. Lett. 26, 525.
Miura, A., 1987. Simulation of Kelvin-Helmholtz instability at the magnetosphere boundary. J. Geophys. Res., 92, 3195-3206.
Morita, Y., Ishikawa, H., 1977. On recent measurements of electric parameters and aerosols in the oceanic atmosphere, in Electrical Processes in Atmosphere, edited by H. Dolezalek and R. Reiter, pp. 126-130, Steinkopff, Darmstadt, Germany.
Muhleisen, R., 1977. The global circuit and its parameters, in Electrical processes in Atmosphere, ed. H. Dolezalek and R. Reiter, Steinkopff, Darmstadt, 467-476.
Nagano, I., Yagitani, S., Migamura, K., Makino, S., 2003. Full wave analysis of elves created by lightning-generated electromagnetic pulses. J. Atmos. Solar Terr. Phys., 65, 615-625.
Neubert, T., 2003. On sprites and their exotic kin. Nature 300, 747-748.
Ney, E.P., 1959. Cosmic radiation and the weather. Nature 183, 451-452.
Nickolaenko, A.P., Hayakawa, M., Hobara, Y., 1996. Temporal variation of the global lightning activity deduced from Schumann resonance. J. Atmos. Terr. Phys. 58, 1699-1709.
Ogawa, T. 1985. Fair-weather electricity. J. Geophys. Res. 90, 5951-5960.
Owon, C. J., Cowley, S. W. H., 1991. Heikkila mechanism for impulsive plasma transport through the magnetopause a re-examination. J. Geophys. Res., 96, 5575-5574.
Parkinson, W.C., Torrenson, O.W., 1931. The diurnal variation the electrical potential of the atmosphere over oceans. Compt. Rend de l'Assemblee de Stockholm, 1930; IUGG, Sect. Terrest. Magn. Electr. Bull. 8, 340 -345.
Pasko, V.P., 2003. Electric jet. Nature, 423, 927-929.
Pasko, V.P., George, J.J., 2002. Three-dimensional modeling of blue jets and blue starters. J. Geophys. Res. 107, 1458, doi:10.1029/2002JA009473.
Pasko, V.P., Inan, U.S., Bell, T.F., Taranenko, Y.N., 1997. Sprites produced by quasi-electrostatic heating and ionization in the lower ionosphere. J. Geophys. Res. 102, 4529-4561.
Pasko V. P., Stanley, M. A., Mathews, J. D., Inan U. S., Wood, T. G., 2002. Electrical discharge from a thundercloud top to the ionosphere. Nature, 416, 152-154, 2002.
Pawar, S.D., Kamra, A.K., 2004. Evaluation of lightning and the possible initiation triggering of lightning discharges by the lower positive charge center in an isolated thundercloud in the tropics. J. Geophys. Res. 109, D02205,doi1029/2003JD003735.
Price, C., 1993. Global surface temperature and the atmospheric electric circuit. Geophys. Res. Lett. 20, 1363-1366.

Price, C., 1994. Lightning, atmospheric electricity and climate change. In: Barron, E.J. (Eds), Fifth Symposium on global change studies and the Symposium on global electrical circuit, global change and the meteorological applications of lightning information. Am. Met. Soc., Boston, MA. p.273.

Price, C., 2000. Evidence for a link between global lightning activity and upper tropospheric water vapour. Nature 406, 290-293.

Price, C., Rind, D., 1992. A simple lightning parameterization for calculating global lighting distributions. Geophys. Res. 97, 9919-9933.

Price, C., Rind, D., 1994. Possible implication of global climate change and global lightning distributions and frequencies. J. Geophys. Res. 99, 10823-10831.

Price, C., Asfur, M., Lyons, W., Nelson, T., 2002. An improved ELF/VLF method for globally geolocating sprites producing lightning. Geophys. Res. Lett. 29, No. 3, 10.1029/2002/GL013519.

Pudovkin, M.I., Babushkina, S.V., 1992. Atmospheric transparency variations associated with geomagnetic disturbances. J. Atmos. Terr. Phys. 54, 1135-1138.

Rastogi, R.G., Patel, V.L., 1975. Effect of interplanetary magnetic field on ionosphere over the magnetic equator. Proc. Indian Acad. Sci. (Earth Planet. Sci.) 82, 121-141.

Reeve, N., Toumi, R., 1999. Lightning activity as an indicator of climate change. Quart. J. Royal Soc. 125, 893-903.

Reid, G.C., 1986. Electrical structure of the middle-atmosphere, in Study in Geophysics: the earth's electrical environment. Eds. Krider, E.P., Roble, R.G., pp 183-194, National Academy Press, Washington, DC.

Reiter, R.,1972. Case study concerning the impact of solar activity upon potential gradient and air earth current in low troposphere. Pure Appl. Geophys. 94, 218-225.

Retails, D., Retails, A., 1998. Effects of air pollution and wind on the large-ion concentration in the air above Athens. J. Geophys. Res. 103, 13927-13932.

Richmond, A.D., 1976. Electric field in the ionospheric and plasmaspheric on quiet days. J. Geophys. Res. 81, 1447-1450.

Richmond, A.D., 1986. Upper-atmospheric electric field sources. In: Study in Geophysics-The Earth's electrical Environment, National Academy Press, Washington, D.C., 195-205.

Roble, R. G., 1985. On solar-terrestrial relation in atmospheric electricity. J. Geophys. Res. 90, 6000-6012.

Roble, R.G., 1991. On modeling component processes in the earth's global electric circuit. J. Atmos. Solar-Terr. Phys., 53, 831-847.

Roble, R.G., Tzur, I., 1986. The global atmospheric electrical circuit. In Study in Geophysics-The Earth's electrical Environment, National Academy Press, Washington, D. C., pp. 206-231.

Rodger, C. J., 1999. Red sprites, upward lightning and VLF Perturbations. Rev. Geophys. 37, 317-336.

Rosenfeld, D., Lensky, I.M., 1998. Space borne-sensed insights into precipitation formation processes in maritime and continental clouds. Bull. Am. Meteorol. Soc. 79, 2457-2476.

Rosenfeld, D., Woodley, W.L., 2003. Closing the 5-year circle: from cloud seeding to space and back to climate change through precipitation physics. In: Tao, W.K., Adler R. (Eds.), Cloud Systems, Hurricanes, and the Tropical Rainfall Measuring.

Rowe, J.F.Jr., Mathews, J.D., 1973. Low-latitude nighttime E region conductivities. J. Geophys. Res. 78, 7461-7470.

Rowland, H. L., 1998. Theories and simulations of elves, sprite and blue jets. J. Atmos. Solar Terr. Phys. 60, 831-844.

Ruhnke, L.H., 1969. Area averaging of atmospheric current, J. Geomag. Geoelectr. 21, 453-462.

Russell, C. T., Fleishman, M., 2002. Joint control of region-2 field aligned current by the east-west component of the interplanetary electric field and polar cap illumination. J. Atmos. Solar Terr. Phys. 64, 1803-1808.

Rycroft, M.J., 1994. Interactions between whistler-mode waves and energetic electrons in the coupled system formed by the magnetosphere, ionosphere and atmosphere. J. Atmos. Terr. Phys. 53, 849-858.

Rycroft, M.J., 2006. Electrical processes coupling the atmosphere and ionosphere: an overview. J.Atmos. Solar-Terr. Phys. 68, 445-456.

Rycroft, M. J., Cho, M., 1998. Modeling electric and magnetic fields due to thunderclouds and lightning from cloud-tops to the ionosphere. J. Atmos. Solar Terrs. Phys. 60, 889-893, 1998.

Rycroft, M.J., Fullekrug, M., 2004. The initiation and evaluation of SPECIAL. J. Atmos. Solar –Terr. Phys. 66, 1103-1113.

Rycroft, M.J., Israelsson, S., Price, C., 2000. The global atmospheric electric circuit, solar activity and climate change. J. Atmos. Solar Terrs. Phys. 62, 1563-1576.

Sapkota, B.K., Varshneya, N.C., 1990. On the global atmospheric electric circuit. J. Atmos. Terr. Phys. 52, 1-20.

Satori, G., 1996. Monitoring Schumann resonance –II. Daily and seasonal frequency variations. J. Atmos. Terr. Phys. 58, 1483-1488.

Satori, G., Zieger, B., 1999. El Nino related meridional oscillation of global lightning activity. Geophys. Res. Lett. 26, 1365-1368.

Schindler, K., 1979. On the role of irregularities in plasma enerty into the magnetosphere. J. Geophys. Res., 84, 7257-7266.

Schlegel, K., Fullekrug, M., 1999. Schumann resonance parameter changes during high energy particle precipitation. J. Geophys. Res. 104, 10111-10118.

Schlegel, K., Diendorfer, G., Thern, S., Schmidt, M., 2001. Thunderstorms, lightning and solar activity-Middle Europe. J. Atmos. Solar Terr. Phys. 63, 1705-1713.

Sentman, D D., 1995. Schuman resonance. In: Volland, H. (Ed.) hand book of Atmospheric Electrodynamics, Vol. 1 CRC Press, Boca Raton, FL, USA.

Sentman, D.D., Fraser, B.J., 1999. Simultaneous observations at Schumann resonance in California and Australia: Evidence for intensity modulation by the local height of the D-region. J. Geophys. Res. 96, 15973-15984.

Sentman, D.D., Heavner, M.J., Baker, D.N., Cayton, T.E., Fraser, B.J., 1996. Effects of solar storm on the Schumann resonance in late 1989, Paper presented at 10[th] Annual Conference on Atmospheric Electricity. Soc. of Atmos. Electr. of Japan, Osaka, Japan.

Singh, D.K., Singh, R.P., Kamra, A.K. 2004. The electrical environment of the Earth's atmosphere: A review. Space Sci. Rev. 113, 375-408.

Singh, R.P., Patel, R.P., Singh, Ashok, K., Das, I.M.L., 2002. Lightning generated ELF, VLF, Optical waves and their diagnostic features, Ind. J. Phys., 76B, 1-15.

Siingh, Devendraa, Singh, R.P., Kamra, A.K., Gupta, P.N., Singh, R., Gopalakrishnan, V., Singh, A.K., 2005. Review of electromagnetic coupling between the Earth's atmosphere and the space environment. J. Atmos. SolarTerr. Phys. 67, 637-658.

Somayajulu, V.V., Reddy, C.A., Viswanathan, K.S., 1985. Simultaneous electric field changes in the equatorial electrojet in phase with polar caps latitude changes during a magnetic storm. Geophys. Res. Lett. 12, 473-475.

Spiro, R. W., Wolf, R. A. 1984. Electrodynamics of convection in the inner magnetosphere, In Magnetospheric current (ed. Poterma, T. A.) (American Geophysics Union, Washington, D.C.) pp 247-259.

Steiger, S.M., Orville, R.E., 2003. Cloud-to-ground lightning enhancement over southern hemisphere. Geophys. Res. Lett. 30, 1975, 10.2029/2003GL017723.

Strangeways, R. J., Raeder, J., 2001. On the transition from collision to collisional magnetohydrodynamics. J. Geophys. Res., 106, 1955-1960.

Stringfellow, M.F., 1974. Lightning incidence in Britain and the solar cycle. Nature. 249, 332-333.

Su, H. T., Hsu, R.R., Chen, A.B., Wang, Y.C., Hsiao, W.S., Lai, W.C., Lee, L.C., Sato, M., Fukunishi, H., 2003. Gigantic jets between a thundercloud and the ionosphere. Nature 423, 974-976.

Svensmark, H, Friss-Christensen, E., 1997. Variation of cosmic ray flux and global cloud coverage-a missing link in solar climate relation. J. Atmos. Solar-Terr. Phys. 60, 667-973.

Takeda, M., Maeda, H., 1980. Three dimensional structure of ionospheric currents-1. Currents caused by diurnal tidal winds. J. Geophys. Res. 85, 6895-6899.

Taranenko, Y.N., Inan, U.S., Bell, T.F., 1993. Interaction with the lower ionosphere of electromagnetic pulses from lightning, heating, attachment and ionization. Geophys. Res. Lett. 20, 1539-1542.

Tinsley, B. A., 1996. Correlations of atmospheric dynamics with solar-wind-induced changes of air-Earth current density into cloud tops. J. Geophy. Res. 101, 29701-29714.

Tinsley, B.A., 2000. Influence of the solar wind on the global electric circuit, and inferred effects on cloud microphysics, temperature, and dynamics of the troposphere. Space Sci. Rev. 94, 231-258.

Tinsley, B.A., Heelis, R. A., 1993. Correlations of atmospheric dynamics with solar activity. Evidence for a connection via the solar wind, atmospheric electricity, and cloud microphysics. J. Geophys. Res. 98, 10375-10384.

Tinsley, B.A., Yu, F., 2003. Atmospheric ionization and clouds as links between solar activity and climate. In: Pap, J. et al. (Eds.), Solar variability and its effects on the Earth's atmospheric and climate system. American Geophysical Union Press, Washington, D.C., pp. 19.

Tinsley, B.A., Zhou, L., Plemons A., 2006. Changes in scavenging of particles by droplets due to weak electrification in clouds. Atmos. Res. 79 266- 295.

Tripathi, S.N., 2000. Removal of charged aerosols. Ph.D. Thesis. The University of Reading, UK.

Tripathi, S.N., Harrison, R.G., 2001. Scavenging of electrified radioactive aerosols. Atmos. Enviorn. 35(33), 5817-5821.

Tripathi, S.N., Harrison, R.G., 2002. Enhancement of contact nucleation by scavenging of charged aerosol. Atmos. Res. 62 57-70

Tsurutani, B.T., and Gonzalez, W.D., 1995. The efficiency of 'viscous interaction' between the solar wind and magnetosphere during intense northward IMF event. Grophys. Res. Lett. 22, 663-666.

Turco, R.P., Zhao, L.X., Yu, F., 1998. A new source of tropospheric aerosol: Ion-ion recombination. Geophys. Res. Lett. 25, 635-638.

Volland, H., 1987. Electromagnetic coupling between lower and upper atmosphere. Physica Scripta T18, 289-297.

Weiss, L. A., Reiff, P. H., Mosses, J. J., Moore, B. D., 1992, Eur. Space Agency Spec. Publ., ESA-SP 335, 309.

Wescott, E.N., Stenback-Mielsen, H.C., Sentman, D.D., Heavner, M.J., Moudry, D.R., Sabbas, F.T.S., 2001. Triangulation of sprites, associated halos and their possible relation to causative lightning and micrometeors. J. Geophy. Res. 106, 10467-10478.

Whipple, F.J.W., Scrase, F.J., 1936. Point discharge in the electric field of the Earth. Geophys. Mimoirs (London), VIII (68), 20.

Wilcox, J.M., Schrrer, P.H., Svalgaard, L., Roberts, W.O., Olson, R.H., Jenne, R.L., 1974. Influence of solar magnetic sector structure on terrestrial atmospheric vorticity. J. Atmos. Sci. 31, 581-591.

Williams, E. R., 1992. The Schumann Resonance: a global thermometer. Science 256, 1184-1187.

Williams, E.R., 1994. Global circuit response to seasonal variations in global surface air temperature. Mon. Weather Rev. 122, 1917-1929.

Williams, E. R., 2005. Lightning and climate: A review. Atmos. Res. 76, 272-287.

Williams, E. R., Heckman, S.J., 1993. The local diurnal variation of cloud electrification and the global diurnal variation of negative charge on the Earth. J. Geophys. Res. 98, 5221-5234.

Williams, E.R., Rothkin, K., Stevenson, D., Boccippio, D., 2000. Global lightning variations caused by changes in thunderstorm flash rate and by changes in the number of thunderstorms. J. Appl. Met. (TRMM Special Issue) 39, 2223-2248.

Williams, E. R., et al., 2002. Contrasting convective regions over the Amazon: implication for cloud electrification. J. Geophys. Res. 107, D20.8082doi10.1029/Jd000380.

Williams, E.R., Mushtak, V.C., Rosenfeld, D., Goodman, S.J., Boccippio, D.J., 2005. Thermodynamic conditions favorable to superlative thunderstorm up draft, mixed phase microphysics and lightning flash rate. Atmos. Res. 76, 288-306.

Wood, C.A., 1951. Unusual Lightning. Weather 5, 230.

**Caption to the Figures:**

Fig. 1   Schematic diagram of various electrical processes in the global electric circuit (Roble and Tzur, 1986). The vector **B** shows the direction of the Earth's geomagnetic field, and arrows show the direction of the current flow in the regions of the tropospheric, ionospheric and magnetospheric generators.

Fig.2   Annual diurnal variation of the potential gradient measured on the surface of the Oceans ( Parkinson and Torrenson, 1931; Whipple and Scrase, 1936).

Fig. 3   Lightning related transient optical emissions in atmosphere (stratosphere and mesosphere): sprites, jets, elves and gigantic (Neubert, 2003; Pasko, 2003).

Fig. 4   Flow chart describing the observed correlations and hypothetical mechanisms between different solar activity and tropical thunderstorms on the global electric circuit, various processes of cloud microphysics and weather and climate. (Tinsely, 2000).

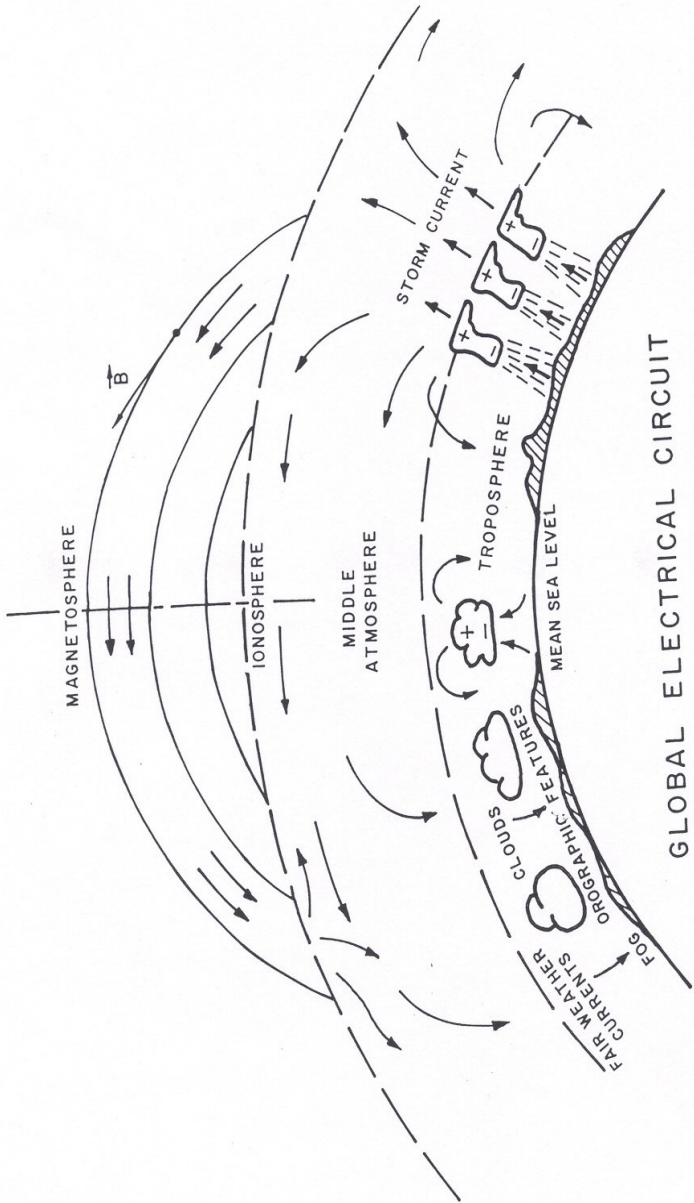

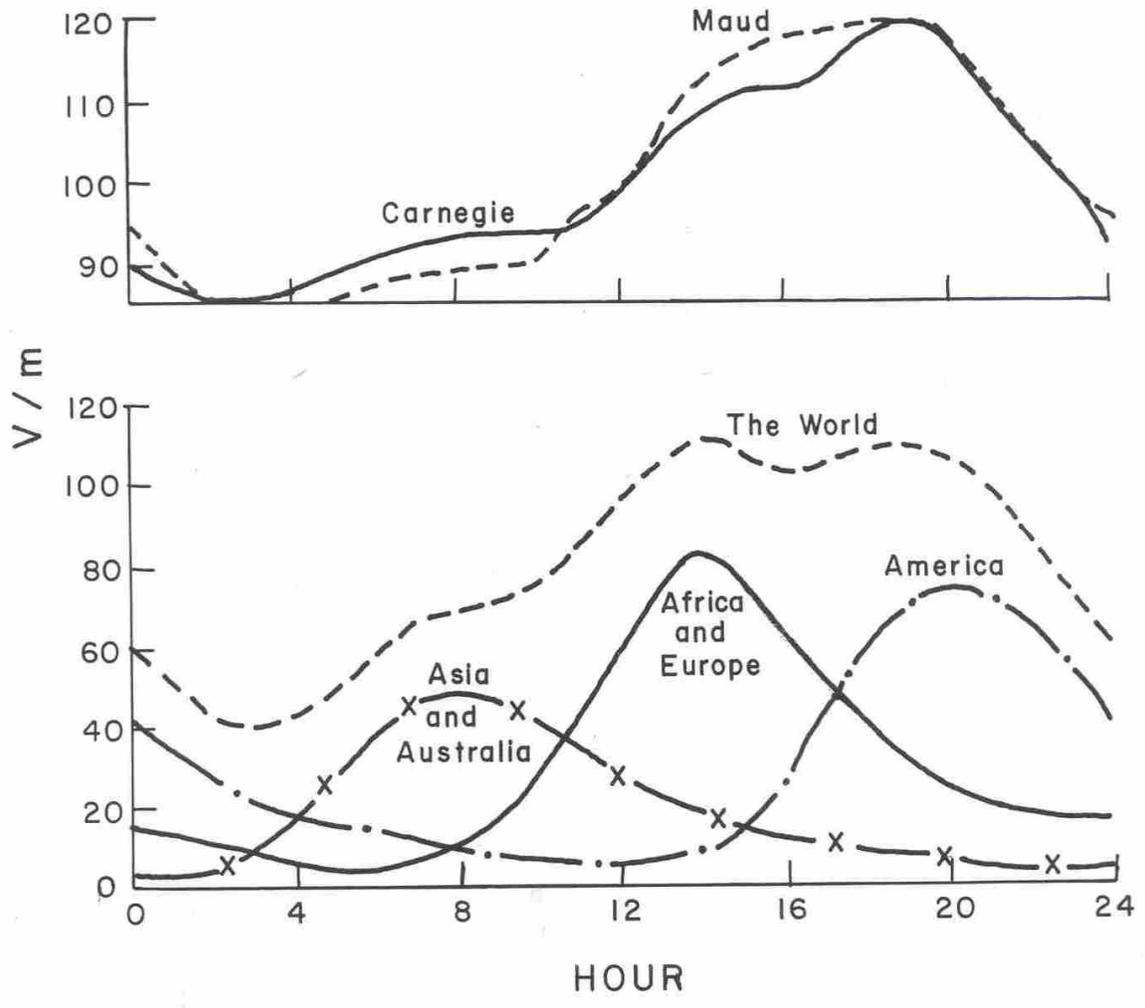

Fig.2

(a)

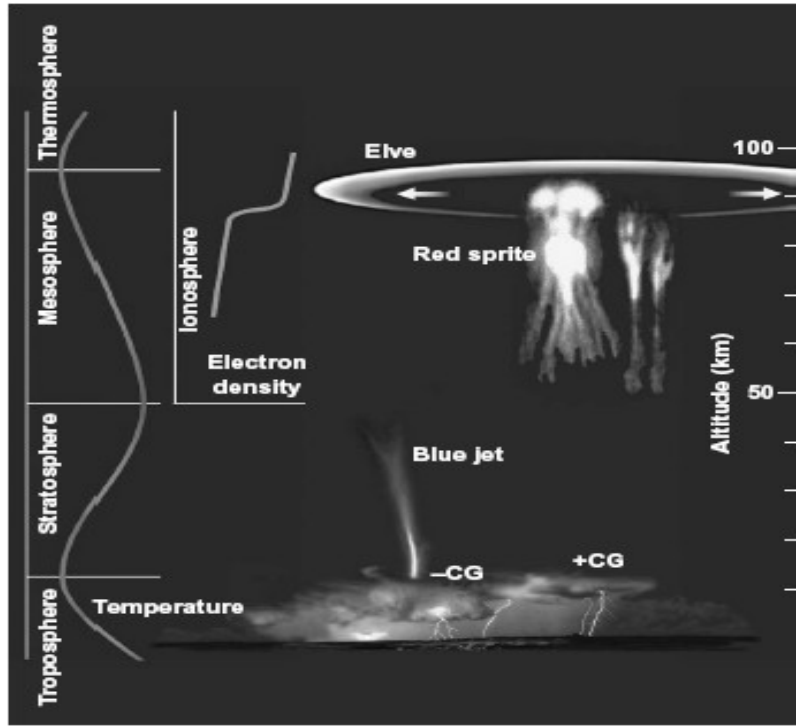

(b)

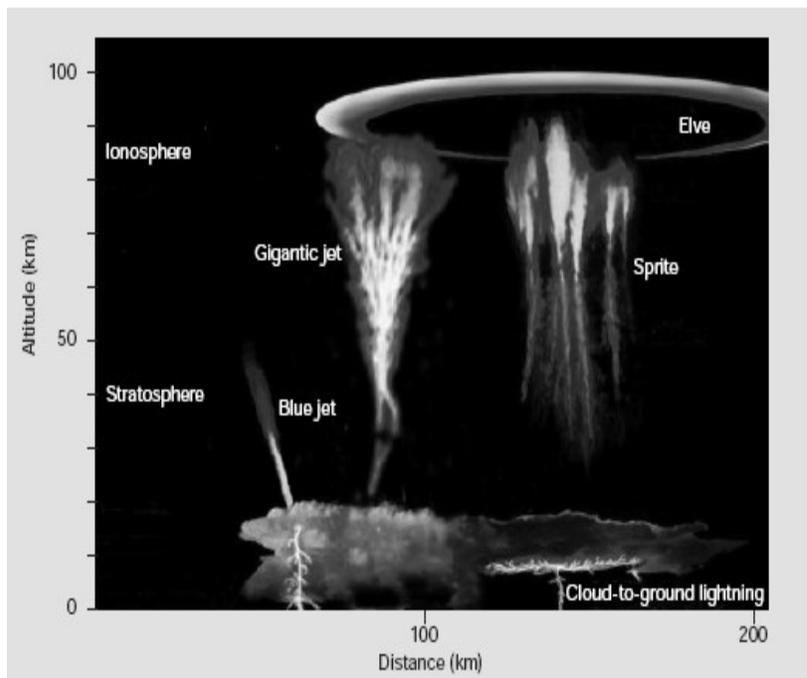

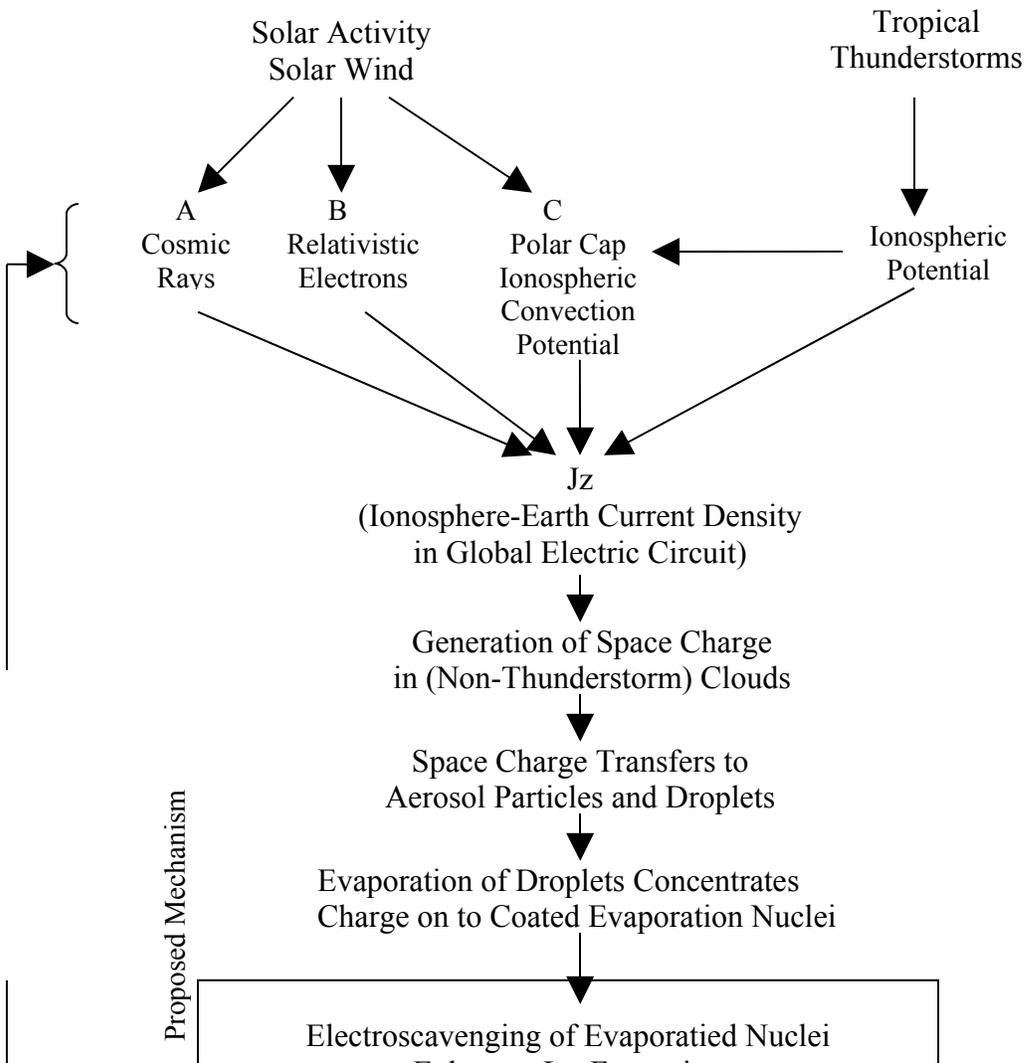

Fig. 4

41